\documentclass[aps,prd,superscriptaddress,amsmath,groupedaddress,reprint,showpacs,10pt,nofootinbib]{revtex4-1}
\usepackage{graphicx}  %
\usepackage{amsmath}
\usepackage{bm}  %
\usepackage{ulem}  %

\newcommand{\bea}{\begin{eqnarray}}
\newcommand{\eea}{\end{eqnarray}}
\newcommand{\beq}{\begin{equation}}
\newcommand{\eeq}{\end{equation}}
\newcommand{\bqa}{\begin{eqnarray}}
\newcommand{\eqa}{\end{eqnarray}}

\def\mqo2{{\!\!\!}}
\renewcommand{\Im}{{\rm Im\,}}

\newcommand{\pv}{\bm{p}}
\newcommand{\qv}{\bm{q}}
\newcommand{\psid}{\psi^\dagger}

\newcommand{\bra}[1]{\langle #1 |}
\newcommand{\ket}[1]{| #1 \rangle}

\newcommand{\Lag}{\mathcal{L}}
\newcommand{\Dv}{\bm{D}}
\newcommand{\rv}{\bm{r}}
\newcommand{\eb}{{\bar e}}
\newcommand{\sigmav}{\boldsymbol{\sigma}}
\newcommand{\nablav}{\boldsymbol{\nabla}}
\newcommand{\eff}{\mathrm{eff}}

\newcommand{\Tr}{\mathrm{Tr}}

\newcommand{\rhoh}{\hat\rho}

\newcommand{\deep}{\mathrm{deep}}

\begin{document}

\title{
Open Effective Field Theories\\
from Deeply Inelastic Reactions}
\author{Eric Braaten}
\affiliation{Department of Physics,
         The Ohio State University, Columbus, OH\ 43210, USA}
\author{H.-W. Hammer}
\affiliation{Institut f\"ur Kernphysik, Technische Universit\"at Darmstadt,
64289\ Darmstadt, Germany}
\affiliation{ExtreMe Matter Institute EMMI, GSI Helmholtzzentrum f\"ur
Schwerionenforschung, 64291\ Darmstadt, Germany}
\author{G.~Peter Lepage}
\affiliation{Laboratory for Elementary Particle Physics, Cornell University,
Ithaca, New York\ 14583, USA}
\date{\today}

\begin{abstract}
Effective field theories have often been applied to systems
with deeply inelastic reactions that produce particles
with large momenta outside the domain of validity of the effective theory.
The effects of the deeply inelastic reactions have been taken into account in
previous work
by adding local anti-Hermitian terms to the effective Hamiltonian.
Here we show that when multi-particle systems are considered,
an additional modification is required in equations governing the density matrix.
We define an effective density matrix by tracing over the states
containing high-momentum particles, and show that it satisfies a Lindblad
equation, with local Lindblad operators
determined by the anti-Hermitian terms in the effective Hamiltonian density.
\end{abstract}

\maketitle

\section{Introduction}
\label{sec:introduction}

Elementary particles are accurately described by local quantum field theories.
Some  of the most fascinating phenomena in condensed matter physics  and in atomic physics
can also  be described by local quantum field theories.
The modern understanding of local quantum field theories is based largely on {\it effective field theory}
\cite{Lepage:1989hf,Georgi:1994qn,Kaplan:1995uv,Manohar:1996cq,Shankar:1996vk,Lepage:1997cs,Burgess:2007pt}.
This approach provides physical interpretations of the mathematical singularities
that are ubiquitous in local quantum field theories.
Effective field theory  provides a systematic framework for
quantifying the effects of higher-momentum physics that can not be described explicitly
within a low-energy effective theory.
Effective field theory also provides a systematic framework for developing low-energy approximations
for phenomena that are described by a local quantum field theory.

One area where effective field theories have proven useful is in the analysis
of the impact of high-momentum decays on nonrelativistic field theories.
Examples include positronium decay into photons,
analyzed in nonrelativistic QED~\cite{Labelle:1993tq,Hill:2000qi},
and the decays of quarkonium states into gluons,
analyzed in nonrelativistic QCD~\cite{Bodwin:1994jh}.
Processes such as these,
whose final-state particles have much larger three-momenta than the inital-state
particles, are mimicked in nonrelativistic effective field theories by local non-Hermitian
corrections to the effective Lagrangian.

Previous analyses
have focused on systems consisting of a single atom or a single meson, where the
treatment of the non-Hermitian corrections is straightforward. In this paper,
we extend this earlier work to include multi-particle systems.
We show that the non-Hermitian terms in the effective Lagrangian lead to
modifications in the evolution equation of the effective density matrix
that describes multi-particle systems. In particular, we show that the
effective density matrix satisfies a
Lindblad equation~\cite{Lindblad:1975ef,Gorini:1975nb,Preskill}. This is true
provided the decay products escape from the system 
or otherwise decouple, so they cannot influence it later.

In Section~\ref{sec:effective-theories}, we review the ideas behind
effective field theories and their use for deeply inelastic processes.
We then outline how these ideas must be adapted for use in multi-particle
systems and the role played by the Lindblad equation. In Section~\ref{sec:example},
we show explicitly how the Lindblad equation emerges from a perturbative analysis of a
simple model. Finally in Section~\ref{sec:conclusions}
we summarize our results and discuss possible applications. 
\section{Effective Theories and Deeply Inelastic Processes}
\label{sec:effective-theories}

\subsection{Two Types of Locality}

\begin{figure}
\centerline{ \includegraphics*[width=8cm,clip=true]{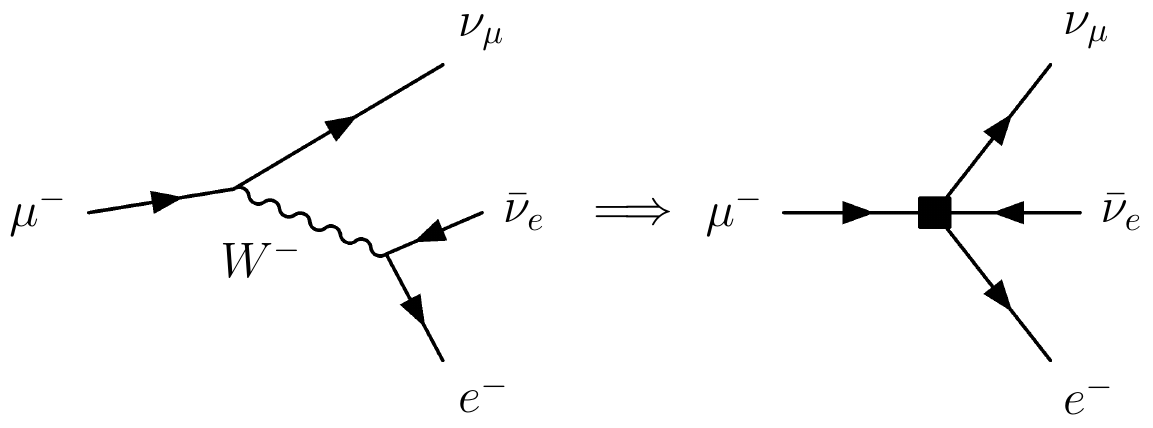} }
\caption{In the amplitude for muon decay,  
the $W$ can propagate only over short distances. Its exchange can therefore
be approximated by a contact interaction.}
\label{fig:MuonDecayAmplitude}
\end{figure}

An effective field theory is obtained by removing ({\it integrating out})
states from a field theory. The simplest applications involve removing
very massive particles. A muon, for example, decays into a
$\nu_\mu$~neutrino and a $W$~boson. The $W$~is almost a thousand
times more massive than the muon and so is highly virtual.
It decays almost immediately ($\Delta t\approx 1/M_W$) into an
electron and the antineutrino~$\bar \nu_e$.
This process is very accurately
modeled by the Fermi interaction (Fig.~\ref{fig:MuonDecayAmplitude}), where the decay occurs at a
point rather than spread over space-time distances of order~$1/M_W$:
    \begin{equation}
    \frac{G_F}{\sqrt{2}}
    \bar\nu_\mu\gamma_\alpha (1-\gamma_5)\mu \,\,
    \bar e\gamma^\alpha (1-\gamma_5) \nu_e.
\end{equation}
We have integrated the $W$ out of the theory.

The Fermi interaction is the leading term in a series of
local operators that can be used to mimic the  decay process
to arbitrary precision. This series is obtained at tree-level by
Taylor expanding the $W$~propagator $1/(q_W^2 - M_W^2)$ in powers
of $q_W^2/m_W^2$ to obtain
\begin{equation}
   \delta\Lag_\eff = \frac{G_F}{\sqrt{2}} \sum_{n=0}^\infty
    \bar\nu_\mu\gamma_\alpha (1-\gamma_5)\mu
    \left(\frac{-\partial^2}{M_W^2}\right)^n
    \bar{e}\gamma^\alpha (1-\gamma_5) \nu_e.
\end{equation}
In practice only one or two terms in this series are needed to
account for experiment. The individual operators are renormalized
when loop corrections are included, but the effective theory
is still capable of reproducing the original theory to arbitrary
precision provided operators of sufficiently high dimension are
retained. Operators of dimension~$n$ correct the theory at
order $(p/M_W)^{n-4}$, where $p$~is the muon's momentum.

\begin{figure}
\centerline{ \includegraphics*[width=8cm,clip=true]{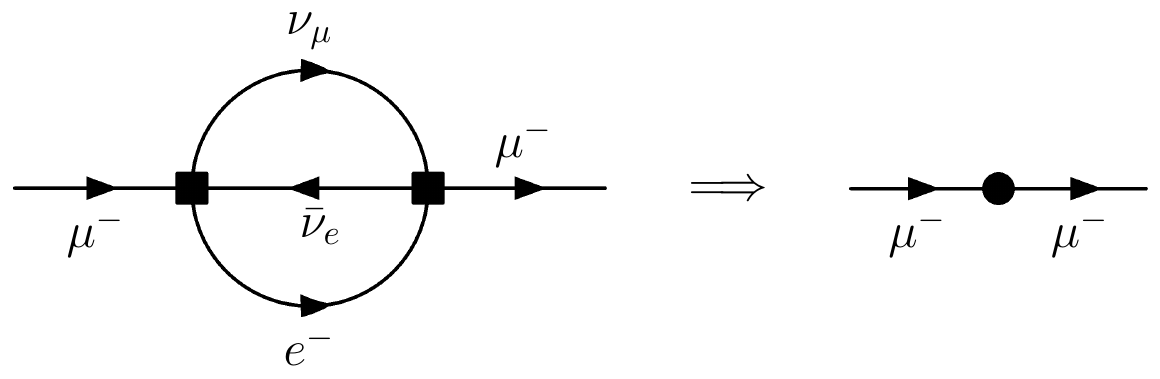} }
\caption{
In the amplitude for $\mu^- \to\nu_\mu e^- \bar \nu_e\to\mu^-$,
the high-momentum intermediate-state particles 
 are created and annihilated in a localized region.
Their effects can therefore be reproduced by local operators.
}
\label{fig:MuonDecayRate}
\end{figure}

A less obvious opportunity to remove states arises when a
particle decays to particles with much smaller masses.
Such a decay is an example of a {\it deeply inelastic
reaction}, where a large fraction of the initial state's rest mass is
converted into large kinetic energies for the final-state particles.
Muon decay is again a good example.
The amplitude for $\mu^- \to \nu_\mu e^- \bar\nu_e \to \mu^-$
on the left side of Fig.~\ref{fig:MuonDecayRate} is analytic in the
muon's three-momentum $\pv$ provided that momentum is nonrelativistic.
This is because the nearest nonanalyticity in the amplitude is at
the threshold energy for the $\nu_\mu e^- \bar\nu_e$~state,
which is effectively zero and  far below the nonrelativistic muon's
energy ($\approx m_\mu$). As a result, we can Taylor expand the amplitude in
powers of~$\pv^2/m_\mu^2$:
\begin{equation}
    T\big[\mu\to \nu_\mu e\bar\nu_e\to\mu\big] =
    T_0 \left(1 + \sum_{n=1}^\infty b_{2n} \left(\frac{-\pv^2}{m_\mu^2}\right)^n\right).
\end{equation}
In practice, only a few terms need to be retained, depending upon
how nonrelativistic the muon is. These corrections can be
incorporated into an effective field theory by discarding the
high-energy $\nu_\mu e^- \bar\nu_e$~states and introducing
new correction terms in the effective theory's Lagrangian:
\begin{equation}
    \delta\Lag_\eff
    = T_0\,\psi_\mu^\dagger
    \left(1 + \sum_{n=1}^\infty b_{2n} \left(\frac{\nabla^{2}}{m_\mu^{2}}\right)^n\right)\,\psi_\mu,
\end{equation}
where $\psi_\mu$ is the (2-component) nonrelativistic muon field.

Here we are particularly interested in the imaginary part of this series,
coming from the muon's deeply inelastic decay reaction.
We can write that part of the effective Lagrangian as
\begin{equation}
    \Lag_\deep
    =  \frac{i}{2}\Gamma_\mu \psi_\mu^\dagger
    \left(1 + \sum_{n=1}^\infty c_{2n} \left(\frac{\nabla^{2}}{m_\mu^{2}}\right)^n\right)\,\psi_\mu,
\end{equation} 
where $\Gamma_\mu$ is the muon's decay rate to~$\nu_\mu e^- \bar\nu_e$.
(The subscript ``deep'' stands for ``deeply inelastic reactions''.)
These terms mimic the effects of muon decay in the effective theory.

Given our first example, it seems nonintuitive
that the effects of a decay to on-shell particles
can be mimicked by local operators. In fact the decay process is quite
local. This is because the location of the decay can be reconstructed
by tracking the decay products back to their origin. The decay products
have relatively short wavelengths of order~$1/m_\mu$ (because of their
high momenta, of order~$m_\mu$), and so
can locate the decay with a resolution of
order~$\Delta x \approx 1/m_\mu$. So the decay is localized over a
region of size~$\Delta x$, which is very small compared to the
wavelength of a nonrelativistic muon ($\gg 1/m_\mu$).

The utility of the effective theory is easily illustrated by
adding QED effects. Corrections
are needed in $\Lag_\deep$ to account for
photons radiated by the decay products (the $W$ or the electron).
Gauge invariance requires the following form:
\begin{align}
   \Lag_\deep = \frac{i}{2}\Gamma_\mu \psi_\mu^\dagger
    &\left\{
    1 + c_2 \frac{\Dv^2}{m_\mu^2} + c_4 \frac{\Dv^4}{m_\mu^4} + \cdots
    \right.
    \nonumber \\
    &\left.
    + f_2 \frac{e\sigmav\cdot\bm{B}}{m_\mu^2}
    + f_3 \frac{e\nablav\cdot\bm{E}}{m_\mu^3}
    + \cdots
    \right\}\psi_\mu,
\end{align} 
where $\Dv$ is the QED gauge-covariant derivative
and $\bm{E}$ and $\bm{B}$ are the electric and magnetic fields.
The coefficients $c_{2n}$ are the same (to leading order)
as in the theory without QED corrections. This formula
shows, for example, that the lifetime of a  $\mu^- e^+$~atom
equals the muon's lifetime up to corrections
of order~$\alpha^2 (m_e/m_\mu)^2\Gamma_\mu$ (due to the $c_2$~term);
in particular there are no  corrections to the binding energy of
order $\alpha^2 (m_e/m_\mu)\Gamma_\mu$~\cite{Czarnecki:1999yj}.

\begin{figure}
\centerline{ \includegraphics*[width=7cm,clip=true]{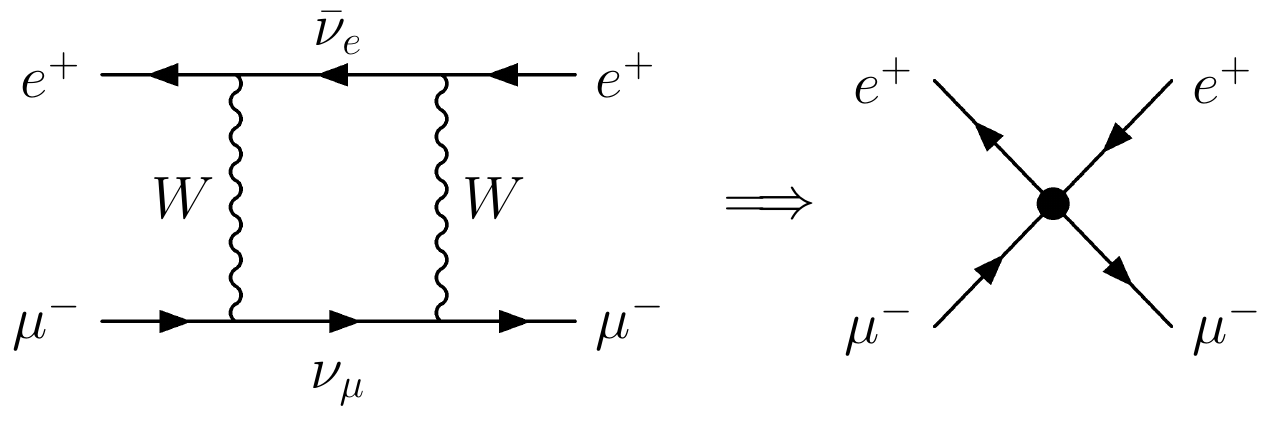} }
\caption{
In the amplitude for $\mu ^- e^+ \to \nu_\mu \bar\nu_e \to \mu^- e^+$,
the high-momentum intermediate-state neutrinos 
are created and annihilated in a localized region.
Their effects can therefore be reproduced by local operators.
}
\label{fig:MuonPositronScat}
\end{figure}

These same ideas apply to deeply inelastic scattering
reactions. For example, the final-state neutrinos in $\mu^- e^+ \to \nu_\mu \bar\nu_e$
have momenta of order~$m_\mu$ when the initial positron and muon are nonrelativistic.
The amplitude for $\mu^-  e^+ \to \nu_\mu \bar\nu_e \to \mu^- e^+$
on the left side of Fig.~\ref{fig:MuonPositronScat}
is analytic in the momenta of $\mu^-$ and $e^+$.
The imaginary part of the amplitude comes from the deeply inelastic scattering reaction.
We can again mimic the effects of the high-momentum final states
using local interactions:
\begin{equation}
    \Lag_\deep = iB_0 \,\psi_\mu^\dagger \psi_\mu \, \psi_\eb^\dagger\psi_\eb
    + iB_1\, \psi_\mu^\dagger\sigmav \psi_\mu \cdot \psi_\eb^\dagger\sigmav\psi_\eb
    + \cdots,
    \label{eq:electron-muon-operators}
\end{equation}
where $B_0$ and~$B_1$ are obtained from the imaginary part of the
amplitude on the left side of Fig.~\ref{fig:MuonPositronScat}.
It is also straightforward to add QED effects here.

\subsection{Multi-particle systems}

The Hamiltonian that follows from the effective theory described in the
previous section has both a Hermitian piece $H_\eff$, associated with
 conventional dynamics, and an anti-Hermitian piece $-iK_\deep$,
coming from the deeply inelastic reactions whose products have been
removed from the theory.\footnote{Contributions to~$K_\deep$ mimic
the anti-Hermitian parts of scattering amplitudes $\bra{b}T\ket{a}$,
where~$\ket{a}$ and~$\ket{b}$ are states in the effective theory
that are connected by intermediate deeply inelastic  reaction channels.
The Hermitian parts of these amplitudes are absorbed into~$H_\eff$.}
 In the case of our nonrelativistic muon,
\begin{subequations}
\begin{align}
    H_\eff &= \int \!\!d^3\rv \,
    \psi^\dagger_\mu \left\{eA_0 - \frac{\Dv^2}{2m_\mu} + \cdots\right\}\psi_\mu,
    \\
    K_\deep &= \frac{1}{2} \Gamma_\mu\int \!\! d^3\rv \,
    \psi^\dagger_\mu \left\{1 + c_2 \frac{\Dv^2}{m_\mu^2} + \cdots\right\}\psi_\mu,
    \label{eq:K-muon-example}
\end{align}
\end{subequations}
where $A_\alpha$ is the photon field.
The leading term in $K_\deep$ is $\frac12 \Gamma_\mu\,\hat N_\mu$,
where $\hat N_\mu$ is the muon number operator:
  \begin{align}
\hat N_\mu = \int d^3\rv \, \psi^\dagger_\mu \psi_\mu .
    \label{eq:Nmuhat}
\end{align}
This Hamiltonian applies to both single-muon and multi-muon systems.

The quantum mechanics of such a theory is
unconventional because probability is not obviously conserved. The Hamiltonian
$H_\eff - iK_\deep$ does {\it not} change the number of muons in a state,
because it commutes with $\hat N_\mu$.
Instead it accounts for the effects of muon decay by reducing
the probability carried by each state:
the norm of a state
that starts with $n$~muons
decays to zero with the decay rate~$n \Gamma_\mu$ (in leading order).
This is the correct
result\,---\,the probability for $n$~muons to still be $n$~muons after
time~$t$ is $\exp(-n\Gamma_\mu t)$.

We typically want more information about where the probability goes.
In the effective theory, an $n$-muon state
evolves into a mixture of
states with $n$, $n-1$, $n-2$\,\ldots\,muons that is most naturally
described by a density matrix.
We can construct an {\it effective density matrix}
$\rhoh_\eff$ from the density matrix $\rhoh$ of the full theory by tracing over 
the deeply inelastic decay products: 
\begin{equation}
\label{eq:rho-eff-defn}
    \rhoh_\eff(t) =  \Tr_\deep \big(\rhoh(t)\big).
\end{equation}
More precisely, we trace out any state containing a particle with momentum
exceeding the ultraviolet cutoff~$\Lambda_\mathrm{UV}$  of the effective field theory.
In the case of our nonrelativistic muon theory, this cutoff is
some fraction of the muon mass $m_\mu$.

The effective density matrix defined by Eq.~(\ref{eq:rho-eff-defn}),
like the  density matrix  of the full theory, 
is Hermitian and non-negative and it has unit trace:  $\Tr(\rhoh_\eff)=1$.
Fourier modes with large frequencies of order $m_\mu$
cannot be described accurately in the effective theory.
Thus the definition of the effective density matrix should also involve a time average 
that eliminates high frequencies.
Such a time average is implicit in Eq.~(\ref{eq:rho-eff-defn}).

The density matrix defined by the partial trace in Eq.~(\ref{eq:rho-eff-defn})
is in general non-Markovian. The time derivative $(d/dt)\rhoh_\eff(t)$ at time $t$
is determined not only by $\rhoh_\eff(t)$ but also by its past history: $\rhoh_\eff(t')$, $t'<t$.
The non-Markovian behavior arises because a high-momentum particle created by a decay 
at time $t'$ can interact with a low-energy particle at a later time $t$.
We make an additional physical assumption that eliminates this possibility.
We assume the high-energy particles from the deeply inelastic reactions interact so weakly 
with the low-energy particles that their subsequent interactions can be ignored. 
This would certainly be the case if the high-momentum particles escape from the system.
Given this assumption, the effective density matrix should be Markovian.

Given that $\rhoh_\eff$ is Markovian,
we might naively expect  its time evolution equation  to be
  \begin{align}
 i\frac{d\ }{dt}\rhoh_\eff \overset{?}{=}   \big[H_\eff, \rhoh_\eff\big]
    - i \big\{K_\deep,\rhoh_\eff\big\},
\end{align}
but  this equation does {\it not}
conserve the total probability~$\Tr(\rhoh_\eff)$.
The correct evolution equation has the structure  of 
the Lindblad equation~\cite{Lindblad:1975ef,Gorini:1975nb}:
if  the local operator $K_\deep$  can be written in the form
\begin{equation}
    K_\deep = \int \!\!d^3\rv\sum_n L_n^\dagger(\rv) L_n(\rv),
    \label{eq:K-structure}
\end{equation}
the Lindblad equation is
  \begin{eqnarray}
  \label{eq:muon-lindblad}
 i\frac{d\ }{dt}\rhoh_\eff &=&   \big[H_\eff, \rhoh_\eff\big]
        - i \big\{K_\deep,\rhoh_\eff\big\}
        \nonumber \\
    &&    +2i  \int \!\!d^3\rv\sum_n L_n(\rv) \, \rhoh_\eff \, L_n^\dagger(\rv).
        \label{eq:drhoeff-Lindblad}
    \end{eqnarray}
The additional term makes $\Tr(\rhoh_\eff)$ time independent, since
the trace of a commutator is zero and the traces of the last two terms 
in~Eq.~(\ref{eq:muon-lindblad}) cancel.
The Lindblad equation is a necessary
consequence of our physical requirements
on the effective density matrix: $\rhoh_\eff$~is 
Hermitian, non-negative, Markovian, and has unit trace.

In the muon decay example, the anti-Hermitian part of the effective
Hamiltonian is given by Eq.~\eqref{eq:K-muon-example}.
There is a single Lindblad operator at leading order, and the Lindblad equation reduces to
\begin{eqnarray}
    i\frac{d}{dt}\rhoh_\eff &=&
    \big[H_\eff, \rhoh_\eff\big]
    - \frac{i}{2} \Gamma_\mu \big\{ \hat N_\mu, \rhoh_\eff\big\}
    \nonumber \\
    &&+ i \Gamma_\mu \int d^3\rv\,\psi_\mu(\rv)\,\rhoh_\eff\,\psi_\mu^\dagger(\rv).
    \label{eq:muon-evol-eq}
\end{eqnarray}

The role of the Lindblad term is easily understood
if we use the evolution equation
to calculate the rate of change of the
probability $P_n(t)$ for finding $n$~muons in the system.
This probability equals the partial trace of $\rhoh_\eff$
over all states $ | X_n \rangle$ that contain $n$ muons:
\begin{equation}
    P_n(t) \equiv \sum_{X_n} \langle X_n | \rhoh_\eff(t) | X_n \rangle.
\end{equation}
The partial trace of the  evolution equation in Eq.~\eqref{eq:muon-evol-eq} gives
\begin{equation}
    \frac{d\ }{dt}  P_n(t)= - n\Gamma_\mu P_n(t) + (n+1)\Gamma_\mu P_{n+1}(t).
\label{eq:dPn(t)}
\end{equation}
The commutator term in
Eq.~(\ref{eq:muon-evol-eq}) does not contribute to the partial trace.
The  anticommutator
term gives $-n\Gamma_\mu P_n$, which is the rate at which
probability leaves the $n$-muon sector because of the decay of a muon. 
The Lindblad term  gives $+(n+1)\Gamma_\mu P_{n+1}$,
which is the rate at which probability enters the $n$-muon sector
from the decay of muons in the $(n+1)$-muon sector.\footnote{This
result follows because $\psi^\dagger(\mathbf{r})$ acting on an
$n$-muon basis state gives $\sqrt{n+1}$ times an $(n+1)$-muon state.}

The Lindblad term in Eq.~(\ref{eq:muon-evol-eq})
is essential to get the correct physical behavior
for the time evolution of the total number of muons.
The expectation value of the muon number, for example, is
\begin{equation}
     N_\mu(t) \equiv {\rm Tr} \big( \hat N_\mu\, \rhoh_\eff(t)\big) = \sum_n n P_n(t).
    \label{eq:Nmu-t}
\end{equation}
We can use Eq.~\eqref{eq:dPn(t)}
to determine the time dependence of $N_\mu(t)$:
    \begin{align}
    \frac{d\ }{dt} N_\mu(t)&= -\Gamma_\mu\Big[
    \sum_n n^2 P_n(t) - \sum_n n(n+1) P_{n+1}(t) \Big].
 \end{align}
After shifting the index of the second term on the right side,
we obtain $(d/dt) N_\mu = - \Gamma_\mu N_\mu$,
which implies that $N_\mu(t) = N_0 \exp(-\Gamma_\mu t)$, as expected.

In order to obtain the Lindblad equation in Eq.~\eqref{eq:drhoeff-Lindblad},
it is essential that $K_\deep$ have the structure shown
in~Eq.~(\ref{eq:K-structure}). This is generally the case
in a nonrelativistic effective field theory.
   In the muon decay example, the operator $K_\deep$ in  Eq.~\eqref{eq:K-muon-example}
can be put into the canonical form in~Eq.~(\ref{eq:K-structure}) by expressing the expansion
inside the braces as the square of the expansion of its square root.
The corresponding corrections to the Lindblad term in Eq.~(\ref{eq:muon-evol-eq})
can be obtained by making the substitution
  \begin{align}
  \psi_\mu(\rv) \longrightarrow
 \left\{1 + c_2 \frac{\Dv^2}{2 m_\mu^2} + \cdots\right\} \psi_\mu(\rv).
\end{align}
More complicated operators in $K_\deep$,
like those that come from the electron-muon terms
in~Eq.~(\ref{eq:electron-muon-operators}), can also be rewritten in
the required form. Such operators have the generic form
\begin{equation}
    K_\deep = \int \!\!d^3\rv \sum_{nm} c_{nm} L_n^\dagger(\rv) L_m(\rv),
    \label{eq:K-structure-gen}
\end{equation}
where the $L_m(\rv)$ are local operators made of 
low-energy fields  and where
$c_{nm}$ is a Hermitian matrix, because $K_\deep$ is Hermitian by definition.
It is also guaranteed to be a positive matrix by the optical theorem:
$-i (T - T^\dagger) = T^\dagger T$.
The double sum is easily rewritten in the canonical
 form of~Eq.~\eqref{eq:K-structure}
 by expanding $c_{nm}$ in terms of outer products
 of its eigenvectors.

\section{A Simple Example}
\label{sec:example}

In this section,  we illustrate how the Lindblad equation for an effective
density matrix emerges naturally from a simple model in perturbation theory.
We consider a theory that describes a nonrelativistic particle of
mass~$M$ (with field~$\psi$) that can decay
into two massless particles (with field~$\phi$).
The  Hamiltonian for the full theory  is
\begin{equation}
\label{eq:hamiltonian}
H = {H}_0^\psi + {H}_0^\phi + H_\mathrm{int},
\end{equation}
where
\begin{subequations}
\begin{eqnarray}
H_0^\psi &=&
 \int_{\bm{r}} \psid(\bm{r}) \left( M - \frac{\nabla^2}{2M} \right) \psi(\bm{r}),
\\
H_0^\phi &=&
\int_{\bm{r}} \tfrac{1}{2}\left(\dot\phi^2 + (\nabla\phi)^2\right),
\\
H_\mathrm{int} &=& \tfrac{1}{2}g \int_{\bm{r}}
    \Big(\psid(\bm{r}) \phi^2(\bm{r}) + \phi^{2}(\bm{r})\psi(\bm{r})\Big).
\label{eq:Hint}
\end{eqnarray}
\end{subequations}
To reduce visual clutter, we have introduced the compact notation
\begin{equation}
     \int_{\bm{r}} \equiv \int\!d^3\rv.
 \end{equation}
The free Hamiltonians describe  particles with on-shell energies
\begin{subequations}
\begin{eqnarray}
    E_{\pv} &=& M + \frac{\pv^2}{2M}, 
 \\
    \omega_{\qv} &=& |\qv|,
\end{eqnarray}
\end{subequations}
respectively.
The interaction  Hamiltonian
allows a $\psi$~particle to decay into a pair of $\phi$~particles.
Our analysis below is simplified if we isolate the part
of the interaction Hamiltonian in Eq.~\eqref{eq:Hint} that causes the deeply inelastic decay:
\begin{equation}
\label{eq:Hdecay}
    H_\deep = \tfrac{1}{2}g 
    \int_{\rv} \Big(\psid(\rv) \tilde\phi^2(\rv)
    + \tilde\phi^{\dagger 2}(\rv) \psi(\rv)\Big),
\end{equation}
where $\phi = \tilde\phi + \tilde\phi^\dagger$ and $\tilde\phi$ is the
annihilation part of $\phi$.

We are interested in systems consisting of nonrelativistic $\psi$~particles,
 whose energies and momenta satisfy
\begin{equation}
    E_{\pv} \approx M \quad\quad |\pv| \ll M.
\end{equation}
The decay products in $\psi\to\phi\phi$ therefore have momenta that are
approximately~$\pm\qv$, where $\qv$ is much larger than the $\psi$'s momentum~$\pv$:
\begin{equation}
    |\qv| \approx M/2 \gg |\pv|.
\end{equation}
We show below how the entire decay process in this limit is effectively local and
instantaneous: it takes place over a spatial region of size
$ \Delta x \sim 1/M$, which is much larger than the  typical length  scale
$1/|\pv|$ associated with a nonrelativistic $\psi$~particle,
and during a time interval  $\Delta t \sim 1/M$,
which is much larger than the  typical time  scale $1/(\pv^2/M)$.
We use
this locality to remove all $\phi$~particles from the theory, creating an
effective theory of unstable $\psi$~particles. We do this first for a single
$\psi$~particle, and then for a system containing multiple $\psi$ particles. Finally,
we show how to adapt these results to a different  model in which $\psi$~particles
are lost through $\psi\psi$~collisions rather than $\psi$ decays. For simplicity,
we assume the coupling~$g$ is small, and we work to leading order in~$g^2$.

\subsection{Locality}
\label{sec:localdecay}

\begin{figure}[t]
\centerline{ \includegraphics*[width=3.16cm,clip=true]{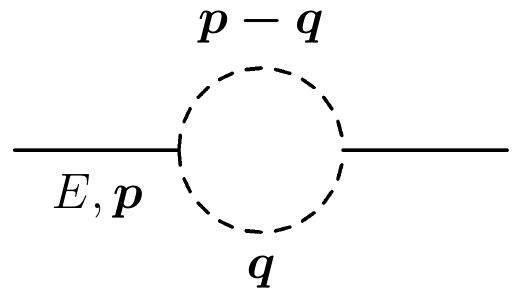} }
\caption{
Self-energy diagram of order $g^2$  for a $\psi$ particle.
The propagators for $\psi$ and $\phi$ are represented by solid lines and dashed lines, respectively.
}
\label{fig:self-energy}
\end{figure}

The leading decay contribution to  the $\psi$ self-energy
comes from the diagram for $\psi\to\phi\phi\to\psi$ in Fig.~\ref{fig:self-energy}:
\begin{equation}
\label{eq:Pidefn}
\begin{split}
    &\tilde\Pi(E,\pv) = \\
    &\quad\quad g^2 \int\!\frac{d^3\qv}{(2\pi)^3}
    \frac{1}{4\omega_{\qv}\omega_{\pv-\qv}}
    \frac{1}{E-\omega_{\pv} - \omega_{\pv-\qv} + i\epsilon},
\end{split}
\end{equation}
where we have used standard time-dependent
perturbation theory to calculate the contribution from $H_\deep$
in Eq.~(\ref{eq:Hdecay})~\cite{Schwartz}.
The integral over~$q$ is dominated by momentum scales of order~$M$ or larger.
This is true as well of its imaginary part evaluated on-shell at $E=E_{\pv}$:
\begin{align}
    &\Im\,\tilde\Pi(E_{\pv},\pv) = \nonumber \\
    &\quad -\tfrac{1}{2}g^2\int\!\frac{d^3\qv}{(2\pi)^3}
    \frac{1}{4\omega_{\qv}\omega_{\pv-\qv}}
    2\pi\delta(E_{\pv} - \omega_{\qv} -\omega_{\pv-\qv}),
\end{align}
since the delta function forces $|\qv|\approx M/2 \gg |\pv|$.
As a result, the distance and time scales that dominate the Fourier transform
of~$\tilde\Pi(E,\pv)$ are of order~$1/M$, and, therefore, the decay process is local
so far as the external (nonrelativistic) $\psi$~particle is concerned.
This also means that we can expand $\tilde\Pi(E_{\pv},\pv)$ in powers of~$\pv^2$:
\begin{equation}
    \tilde\Pi(E_{\pv}, \pv) = \tilde\Pi(M,0) \left(1 + \sum_{n=1}^\infty c_n
    \left(\frac{\pv^2}{M^2}\right)^{n}\right),
\end{equation}
where the coefficients $c_n$ are independent of $\pv$.
We are interested
here in the leading term ($n=0$) in the power series
expansion of~$\tilde\Pi(E_{\pv},\pv)$.  We ignore the remaining terms in what follows.
They are easily included as higher-order corrections to the effective
Hamiltonian, as we discussed in Section~\ref{sec:effective-theories}.

The leading effect of $\tilde\Pi(E,\pv)$ on a single-$\psi$ state is to renormalize
its free Hamiltonian:
\begin{equation}
    H_0^\psi \longrightarrow H_\mathrm{eff} - iK_\deep,
\end{equation}\
where
\begin{subequations}
\begin{eqnarray}
    H_\mathrm{eff} &=& 
\int_{\rv}  \psid(\rv) \bigg( M - \frac{\nabla^2}{2M}+ \mathrm{Re}\,\tilde\Pi(M,0) \bigg)\psi(\rv),
\nonumber\\
\label{Heff-psi}
\\
    K_{\deep} &=& \frac{1}{2} \Gamma \int_{\rv} \psid(\rv)\psi(\rv).
\end{eqnarray}
\end{subequations}
 In the Hermitian part of the effective Hamiltonian, $\mathrm{Re}\,\tilde\Pi(M,0)$
is absorbed into a renormalization of the mass~$M$. 
In the anti-Hermitian part,
$\Gamma$ is the decay rate of a $\psi$ particle:
\begin{equation}
    \Gamma = - 2 \,\mathrm{Im}\,\tilde\Pi(M,0).
\label{Gamma-Pi}
\end{equation}

\begin{figure}[t]
\centerline{ \includegraphics*[width=8cm,clip=true]{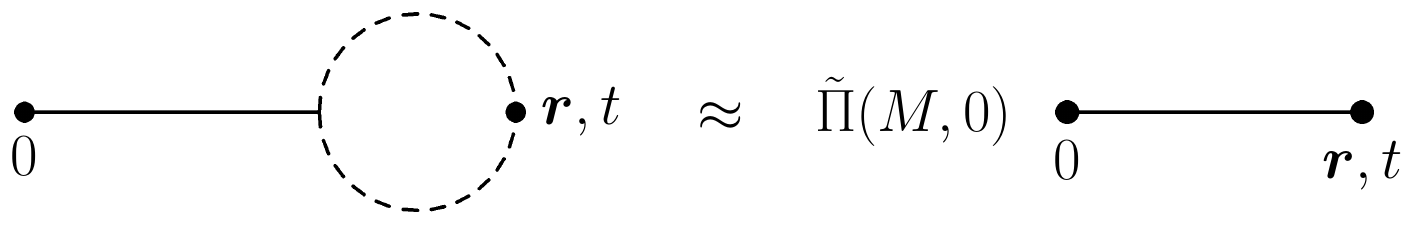} }
\caption{
The operator $\tfrac{1}{2}g\tilde\phi^2(\rv,t)$ can be replaced by
$\tilde\Pi(M,0)\psi(\rv,t)$ in a nonrelativistic correlator.
}
\label{fig:phi2-replacement}
\end{figure}

The locality of the decay process for nonrelativistic momenta, which implies
$ \tilde\Pi(E,\pv)\approx\tilde\Pi(M,0)$,
allows us to simplify correlators that involve $\tilde\phi$~fields
and $\psid$~fields.
For example, in the absence of interactions, the correlator
$\bra{0}\tilde\phi^2(\rv,t)\,\psid(0,0)\ket{0}$ vanishes. With interactions,
we can use locality to replace the $\tilde\phi$~fields by a $\psi$ field,
as illustrated in  Fig.~\ref{fig:phi2-replacement}:
\begin{align}
\label{eq:phi2-substitution}
    \tfrac{1}{2}g \tilde\phi^2(\rv) \longrightarrow
    \tilde\Pi(M,0)\,\psi(\rv,t) 
\end{align}
when the operators are acting to the right on a state in the Fock space of $\psi$.
Similarly, we can replace
\begin{align}
\label{eq:phidag2-substitution}
\tfrac{1}{2}g \tilde\phi^{\dagger 2}(\rv) \longrightarrow
  \tilde\Pi^*(M,0) \,\psid(\rv)
\end{align}
when the operators are acting to the left on a state in the Fock space of $\psi$.
We use these substitutions in the next section.

\subsection{Emergence of the Lindblad equation}
\label{sec:Effrho}

Replacing the free Hamiltonian $H_0^\psi$
by the effective Hamiltonian $H_\mathrm{eff} - iK_\mathrm{dir}$ is all
that is needed
to analyze the impact of the high-momentum decay on single-$\psi$ states.
Analyzing multi-$\psi$ states is more complicated, however, as we discussed
in Section~\ref{sec:effective-theories}:
a system that is described initially by a state with~$n$ $\psi$~particles
evolves into a mixture of states with $n$, $n-1$, $n-2$, \ldots\ $\psi$~particles.
The single-$\psi$ state also evolves into a mixture, but there are only
two states, $n=1$ and~$n=0$, and we don't care about the second one. For
$n>1$, we need the density matrix~$\rhoh(t)$ to track the superposition of states
containing different numbers of~$\psi$~particles over time.
Following Section~\ref{sec:effective-theories}, we
replace this density matrix by
an effective density matrix obtained by tracing over the Hilbert space
of the decay products:
\begin{equation}
    \rhoh_\eff(t) \equiv \mathrm{Tr}_\phi \, \rhoh(t).
\end{equation}
Our goal is to derive an evolution equation for~$\hat\rho_\eff$.

The temporal evolution of the density matrix is given by
\begin{equation}
    i\frac{d\ }{dt} \rhoh =
    \big[H,\rhoh\big],
\end{equation}
where $H$ is the Hamiltonian for the full theory in Eq.~(\ref{eq:hamiltonian}).
We obtain an evolution equation for the effective density matrix by tracing
both sides of this equation over all $\phi$~states:
\begin{equation}
\label{eq:ddtrhoeff}
        i\frac{d\ }{dt} \rhoh_\eff =
   \Tr_\phi \big[H,\rhoh\big].
\end{equation}
The contributions to this equation from the kinetic terms are simple since
\begin{subequations}
\begin{eqnarray}
    \Tr_\phi \big[H_0^\psi, \rhoh \big]
    &=& \big[H_0^\psi, \rhoh_\eff\big],
\\
    \Tr_\phi \big[H_0^\phi, \rhoh \big] &=& 0,
\end{eqnarray}
\end{subequations}
The first equation holds because $H_0^\psi$ does not act on $\phi$~states.
The second equation holds because $H_0^\phi$ depends only on $\phi$ fields.\footnote{
The identity $\Tr_\phi ( \hat A \hat B ) = \Tr_\phi ( \hat B \hat A )$
holds for any operator $\hat A$ constructed out of the field $\phi$
and any operator $ \hat B$.  This can be verified by expressing the partial trace 
as a sum over a complete set of $\phi$ states and inserting a complete set of 
 $\phi$ states between $\hat A$ and $\hat B$.} 
The evolution equation \eqref{eq:ddtrhoeff} reduces to
\begin{equation}
\label{eq:rhoeff-evolution1}
    i\frac{d\ }{dt} \rhoh_\eff = \big[H_0^\psi, \rhoh_\eff\big]
    + \Tr_\phi \big[H_\mathrm{int},\rhoh\big].
\end{equation}

Again we focus on the interaction term in Eq.~\eqref{eq:Hdecay} that causes decays.
The other parts of the interaction term generate additional contributions to the Hermitian
part of the effective Hamiltonian ($H_\eff$), which 
we ignore because they are irrelevant to decays.
The decay interaction contributes four terms
to the right side of the evolution equation (\ref{eq:rhoeff-evolution1}):
\begin{align}
\label{eq:Hdecay-contribution}
     &\Tr_\phi \big[H_\deep,\rhoh\big] \nonumber \\
     &\quad= \tfrac{1}{2}g \int_{\rv} \Tr_\phi \Big(\psid(\rv) \tilde\phi^2(\rv)\, \rhoh +
      \tilde\phi^{\dagger 2}(\rv)\psi(\rv)\, \rhoh\Big) \nonumber \\
     &\quad- \tfrac{1}{2}g \int_{\rv} \Tr_\phi \Big( \rhoh\, \psid(\rv) \tilde\phi^2(\rv)  +
      \rhoh\, \tilde\phi^{\dagger 2}(\rv)\psi(\rv) \Big).
 \end{align}

For simplicity, we consider the initial time when the system consists only of $\psi$ particles.
The action of $\rho(t)$ at this time includes a projection onto the Fock space of $\psi$.
We proceed to examine each of the four terms in Eq.~\eqref{eq:Hdecay-contribution} in turn.
We can use the substitution in Eq.~(\ref{eq:phi2-substitution}) to rewrite the
first trace in Eq.~\eqref{eq:Hdecay-contribution} as:
\begin{equation}
\label{eq:ph2-example}
\tfrac{1}{2}g \Tr_\phi
\Big(\psid(\rv) \tilde\phi^2(\rv)\, \rhoh\Big)
\approx \tilde\Pi(M,0)\, \psid(\rv)\psi(\rv)\,\rhoh_\eff.
\end{equation}
Similarly we can use Eq.~(\ref{eq:phidag2-substitution})
to rewrite the trace in the second decay term as:
\begin{align}
\tfrac{1}{2}g \Tr_\phi
\Big(\tilde\phi^{\dagger 2}(\rv) \psi(\rv)  \,\rhoh\Big) &=
\tfrac{1}{2}g  \Tr_\phi
\Big( \psi(\rv)\,  \rhoh\, \tilde\phi^{\dagger 2}(\rv)\Big)
\nonumber \\
&\approx
\tilde\Pi^*(M,0)\, \psi(\rv)\,\rhoh_\eff\,\psid(\rv).
\end{align}
The traces in the remaining two terms follow the same patterns:
\begin{subequations}
\begin{eqnarray}
\tfrac{1}{2}g \Tr_\phi \big( \rhoh \psid(\rv) \tilde\phi^2(\rv)\big)
&\approx&
\tilde\Pi(M,0)\psi(\rv)\rhoh_\eff\psid(\rv),
\\
\tfrac{1}{2}g \Tr_\phi \big( \rhoh \tilde\phi^{\dagger 2}(\rv)\psi(\rv) \big)
&\approx&
\tilde\Pi^*(M,0)\rhoh_\eff\psid(\rv)\psi(\rv).
\end{eqnarray}
\end{subequations}
Inserting these traces into Eq.~(\ref{eq:Hdecay-contribution}),
we obtain our final result for the evolution equation in Eq.~(\ref{eq:rhoeff-evolution1}):
\begin{eqnarray}
i\frac{d}{dt}\rhoh_\eff = \big[H_\eff, \rhoh_\eff\big]
- \frac{i}{2} \Gamma\int_{\rv}
\Big(\psid(\rv)\psi(\rv)\,\rhoh_\eff
\nonumber \\
 + \rhoh_\eff\,\psid(\rv)\psi(\rv) -2 \psi(\rv)\,\rhoh_\eff \,\psid(\rv) \Big),
\end{eqnarray}
where $H_\eff$ and $\Gamma$ are defined in 
Eqs.~\eqref{Heff-psi} and \eqref{Gamma-Pi}.
This equation has the standard Lindbladian form.
The last term removes $\psi$~particles one at a time
to account for their disappearance due to decays into
pairs of high-momentum $\phi$~particles.

Note that we are making a nontrivial physical assumption about~$\rhoh$
when we use, for example, Eq.~(\ref{eq:phi2-substitution})
to remove $\tilde\phi$~fields
from the effective evolution equation, as in~Eq.~(\ref{eq:ph2-example}). This
substitution is valid provided $\tilde\phi^2$ annihilates $\phi$ particles
coming from the $\psi$~sector of the density matrix (that is from $\psi$~decays).
In principle, it is also possible for~$\tilde\phi^2(\rv)$ to annihilate
$\phi$~particles from the $\phi$-sector of~$\rhoh$. We assume that such contributions
can be ignored because the probability for finding two $\phi$~particles
at the same space-time point is vanishingly small (and therefore
the probability of an inverse
decay, $\phi\phi\to\psi$, is negligible).
This is the case if $\rhoh$ describes a situation in which all
$\phi$~particles are produced by $\psi$~decays and, once produced,
they escape from the system  or otherwise decouple.

\subsection{Inelastic scattering}

\begin{figure}[t]
\centerline{ \includegraphics*[width=3cm,clip=true]{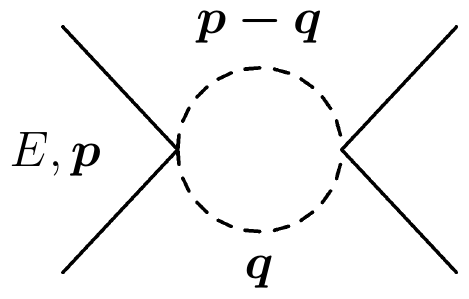} }
\caption{
Diagram of order $g^2$ for $\psi\psi \to \psi\psi$
through an intermediate state with two $\phi$~particles.
}
\label{fig:scattering}
\end{figure}

A variation on our simple model is to replace the decay process
$\psi\to\phi\phi$  by a deeply inelastic scattering process $\psi\psi\to\phi\phi$
as the mechanism by which probability leaks from the $\psi$~sector.
We replace the interaction Hamiltonian in Eq.~\eqref{eq:Hint} by
\begin{align}
    H_\mathrm{int} =
    \tfrac{1}{4}g \int_{\bm{r}}
    \Big(\psi^{\dagger2}(\bm{r})\phi^2(\bm{r})
    + \phi^{2}(\bm{r})\psi^2(\bm{r})\Big).
\end{align}
This interaction term allows the inelastic scattering reaction $\psi\psi\to\phi\phi$,
where now the decay products have approximate momenta $\pm\qv$ with $|\qv|\approx M$.
The leading contribution to the transition amplitude for $\psi\psi \to \psi \psi$
comes from the diagram in Fig.~\ref{fig:scattering}.

The analysis of the effective density matrix for this model,
where the $\phi$~states are
traced out, is almost identical to our decay model analysis above.
Here, in place of Eqs.~(\ref{eq:phi2-substitution})
and~(\ref{eq:phidag2-substitution}), we have substitutions
\begin{subequations}
\begin{eqnarray}
    \tfrac{1}{2}g\, \tilde\phi^2(\bm{r},t) &\longrightarrow& \tfrac{1}{2} \tilde\Pi(2M,0)\,\psi^2(\bm{r},t),
     \\
    \tfrac{1}{2}g\, \tilde\phi^{\dagger2}(\bm{r},t)
     &\longrightarrow& \tfrac{1}{2} \tilde\Pi^*(2M,0)\,\psi^{\dagger2}(\bm{r},t),
\end{eqnarray}
\end{subequations}
where $\tilde\Pi(E,\pv)$ is the same function defined in Eq.~(\ref{eq:Pidefn}).
The new term in the effective Hamiltonian  is an interaction term
instead of a mass term:
\begin{align}
    H_\eff - i K_\deep = H_0^\psi
    + \tfrac{1}{4} \tilde\Pi(2M,0)\,\int_{\bm{r}}\psi^{\dagger2}(\bm{r})\psi^2(\bm{r}).
\end{align}
The anti-Hermitian part of the  effective Hamiltonian
comes from the imaginary part of $\tilde\Pi(2M,0)$, which we denote by $-\Gamma/2$.
The final evolution equation for the effective density matrix is
\begin{eqnarray}
i\frac{d}{dt}\rhoh_\eff = \big[H_\eff, \rhoh_\eff\big]
-\frac{i}{2}\Gamma\int_{\rv}
\Big(\psi^{\dagger 2}(\rv)\psi^2(\rv)\,\rhoh_\eff
\nonumber \\
+ \rhoh_\eff\,\psi^{\dagger 2}(\rv)\psi^2(\rv)
-2 \psi^2(\rv)\,\rhoh_\eff \,\psi^{\dagger 2}(\rv) \Big).
\end{eqnarray}
This equation again has the standard Lindbladian form.
The last term removes $\psi$~particles two at a time
to account for their disappearance due to inelastic scattering into
pairs of high-momentum $\phi$~particles.

\section{Discussion}
\label{sec:conclusions}

The effective Hamiltonian for an effective field theory obtained by integrating out 
high-momentum particles produced by deeply inelastic reactions is local but non-Hermitian.
We have pointed out that states consisting of low-energy particles are naturally described by
an effective density matrix obtained by tracing over states containing
high-momentum particles, as in Eq.~\eqref{eq:rho-eff-defn}.
The time evolution of the effective density matrix is given by the Lindblad equation
in Eq.~\eqref{eq:drhoeff-Lindblad}.
The Lindblad operators $L_n(\rv)$ are local, and they can be deduced from the anti-Hermitian terms 
in the effective Hamiltonian density, which can be expressed in the form in Eq.~\eqref{eq:K-structure}.
The Lindblad terms in the evolution equation are essential to get the 
correct behavior for the time evolution of multiparticle observables,
such as the number of low-energy particles.

The Lindblad equation is familiar in quantum information theory  \cite{Preskill}.
An {\it open quantum system} consists of a {\it subsystem} of interest together with its {\it environment}.
A time evolution equation for the density matrix of the subsystem is called a {\it master equation}.
Under special conditions, 
the master equation has the form of the Lindblad equation \cite{Lindblad:1975ef,Gorini:1975nb}.
These conditions ensure that the autocorrelation function of the
interaction Hamiltonian that connects the subsystem and the environment
decreases to 0 at large times.

An {\it open effective field theory} is an open quantum system in which the subsystem of interest 
is an effective field theory \cite{Grozdanov:2013dba,Burgess:2014eoa}.
Grozdanov and Polonyi have proposed an open effective field theory for the hydrodynamic modes 
of a quantum field theory as a framework for deriving dissipative hydrodynamics \cite{Grozdanov:2013dba}.
Burgess, Holman, Tasinato, and Williams have applied open effective field theory 
to the super-Hubble modes  of primordial quantum fluctuations 
in the early universe \cite{Burgess:2014eoa,Burgess:2015ajz}.
In the stochastic inflation framework, the master equation is  the Lindblad equation.
We have shown that an effective field theory in which deeply inelastic reaction products 
have been integrated out is an open effective field theory.
In this case, the environment consists of the high-momentum particles 
produced by the deeply inelastic reactions.

A heavy quark and heavy antiquark  in the quark-gluon plasma 
can be regarded as an open quantum system in which the heavy quark-antiquark pair 
is the subsystem of interest and the quark-gluon plasma is the environment. 
The quark-gluon plasma can cause the decoherence of the 
heavy quark-antiquark pair and the dissociation of heavy-quarkonium bound states.
A master equation for the heavy quark-antiquark subsystem 
that has the Lindblad form has been derived \cite{Akamatsu:2014qsa}.
This problem could perhaps be formulated in terms of an open effective field theory using 
potential NRQCD \cite{Brambilla:1999xf}.

Ultracold atoms can be described by a local nonrelativistic effective field theory 
for which the coupling constant is  the scattering length \cite{Braaten:2004rn}.
Many loss processes for ultracold atoms involve deeply inelastic reactions.
An important example is three-body recombination,
in which a collision of three low-energy atoms results in the binding of two of the atoms
into a diatomic molecule with a large binding energy.
The Lindblad equation is useful for 
deriving universal relations for the loss rate of ultracold atoms \cite{BHP}.

Open effective field theories from integrating out deeply inelastic reactions
may have other applications in high energy physics.
One particularly interesting application is dark matter.
The deeply inelastic reactions are annihilation collisions of pairs of dark matter particles, which 
produce Standard Model particles that may be observed in indirect detection experiments.
The Lindblad equation could prove to be especially useful if  dark matter particles
have strong self-interactions or if they are in a Bose-Einstein condensate.

\begin{acknowledgments}
E.B. and H.-W.H.\ acknowledge useful discussions with Shina Tan.
The research of E.B.\ was supported in part by the Department of Energy
under grant DE-FG02-05ER15715, by the
National Science Foundation under grant PHY-1310862,
and by the Simons Foundation.
The research of H.-W.H.\ was supported in part
by the BMBF under contract 05P15RDFN1,
and by the Deutsche Forschungsgemeinschaft (SFB 1245).
The research of G.P.L.\ was supported in part by
the National Science Foundation.
\end{acknowledgments}





\begin{thebibliography}{99}

\bibitem{Lepage:1989hf}
  G.~P.~Lepage,
  What is renormalization?,
  in {\it From Actions to Answers (TASI 89)}, eds. T.~DeGrand and D.~Toussaint
  (World Scientific, 1989) [hep-ph/0506330].

\bibitem{Georgi:1994qn}
  H.~Georgi,
  Effective field theory,
  Ann.\ Rev.\ Nucl.\ Part.\ Sci.\  {\bf 43}, 209 (1993).

\bibitem{Kaplan:1995uv}
  D.B.~Kaplan,
  Effective field theories,
  nucl-th/9506035.

\bibitem{Manohar:1996cq}
  A.V.~Manohar,
  Effective field theories,
  Lect.\ Notes Phys.\  {\bf 479}, 311 (1997)
  [hep-ph/9606222].

\bibitem{Shankar:1996vk}
  R.~Shankar,
  Effective field theory in condensed matter physics,
  in {\it Conceptual Foundations of Quantum Field Theory},
  ed.\ T.Y.~Cao (Cambridge University Press, 1999)
[cond-mat/9703210].

\bibitem{Lepage:1997cs}
  G.P.~Lepage,
  How to renormalize the Schrodinger equation, in 
  \textit{Proceedings, 8th Jorge Andre Swieca Summer School on Nuclear Physics},
  ed.~C.A. Bertulani et al. (World Scientific, 1997)
  [nucl-th/9706029].

\bibitem{Burgess:2007pt}
  C.P.~Burgess,
Introduction to Effective Field Theory,
  Ann.\ Rev.\ Nucl.\ Part.\ Sci.\  {\bf 57}, 329 (2007)
  [hep-th/0701053].


\bibitem{Labelle:1993tq}
  P.~Labelle, G.P.~Lepage and U.~Magnea,
Order $m \alpha^8$ contributions to the decay rate of orthopositronium,
  Phys.\ Rev.\ Lett.\  {\bf 72}, 2006 (1994)
  [hep-ph/9310208].

\bibitem{Hill:2000qi}
  R.J.~Hill and G.P.~Lepage,
Order $\alpha^2 \Gamma$ binding effects in orthopositronium decay,
  Phys.\ Rev.\ D {\bf 62}, 111301 (2000)
  [hep-ph/0003277].

\bibitem{Bodwin:1994jh}
  G.T.~Bodwin, E.~Braaten and G.P.~Lepage,
Rigorous QCD analysis of inclusive annihilation and production of heavy quarkonium,
  Phys.\ Rev.\ D {\bf 51}, 1125 (1995)
  [hep-ph/9407339].

\bibitem{Lindblad:1975ef}
  G.~Lindblad,
On the generators of quantum dynamical semigroups,
  Commun.\ Math.\ Phys.\  {\bf 48}, 119 (1976).

\bibitem{Gorini:1975nb}
  V.~Gorini, A.~Kossakowski and E.C.G.~Sudarshan,
Completely positive dynamical semigroups of $N$-level systems,
  J.\ Math.\ Phys.\  {\bf 17}, 821 (1976).

\bibitem{Preskill}
For an accessible discussion and derivation of the Lindblad equation
in a modern context (quantum computing) that is conceptually quite
similar to our problem see:  J.~Preskill,
Lecture Notes for Physics 229: Quantum Information and Computation
(unpublished), Chapter~3.

\bibitem{Czarnecki:1999yj}
  A.~Czarnecki, G.P.~Lepage and W.J.~Marciano,
Muonium decay,
  Phys.\ Rev.\ D {\bf 61}, 073001 (2000)
  [hep-ph/9908439].

\bibitem{Schwartz}
For a simple illustration of time-dependent perturbation theory
applied to quantum field theory, see Chapter~4 in M.~Schwartz, Quantum
Field Theory and the Standard Model (Cambridge University Press, 2014).

\bibitem{Grozdanov:2013dba} 
  S.~Grozdanov and J.~Polonyi,
Viscosity and dissipative hydrodynamics from effective field theory,
  Phys.\ Rev.\ D {\bf 91}, 105031 (2015)
  [arXiv:1305.3670].
  
\bibitem{Burgess:2014eoa}
C.P.~Burgess, R.~Holman, G.~Tasinato and M.~Williams,
EFT beyond the horizon: stochastic inflation and how primordial quantum fluctuations go classical,
  JHEP {\bf 1503}, 090 (2015)
  [arXiv:1408.5002].

\bibitem{Burgess:2015ajz}
  C.P.~Burgess, R.~Holman and G.~Tasinato,
Open EFTs, IR effects and late-time resummations: systematic corrections in stochastic inflation,
  arXiv:1512.00169.

\bibitem{Akamatsu:2014qsa} 
  Y.~Akamatsu,
Heavy quark master equations in the Lindblad form at high temperatures,
  Phys.\ Rev.\ D {\bf 91}, 056002 (2015)
  [arXiv:1403.5783].
  
\bibitem{Brambilla:1999xf} 
  N.~Brambilla, A.~Pineda, J.~Soto and A.~Vairo,
Potential NRQCD: An Effective theory for heavy quarkonium,
  Nucl.\ Phys.\ B {\bf 566}, 275 (2000)
  [hep-ph/9907240].
 
\bibitem{Braaten:2004rn}
  E.~Braaten and H.-W.~Hammer,
Universality in few-body systems with large scattering length,
  Phys.\ Rept.\  {\bf 428}, 259 (2006)
  [cond-mat/0410417].

\bibitem{BHP}
  E.~Braaten, H.-W.~Hammer, and G.P.~Lepage, in preparation.

\end{thebibliography}
\end{document}